\documentclass[final,5p,times,twocolumn]{elsarticle}
\usepackage{textgreek}
\usepackage{verbatim} 
\usepackage{amssymb}
\usepackage{url}
\usepackage{amsmath,amssymb}
\usepackage{mathtools} 
\usepackage[symbol]{footmisc}
\usepackage{subcaption}
\usepackage{capt-of}
\usepackage{bigints}
\usepackage{bm}
\usepackage{xcolor} % For \hl
\usepackage{soul} % For \hl
\usepackage{lipsum}
\def\be{\begin{eqnarray}}
\def\ee{\end{eqnarray}}
\def\nn{\nonumber}

\journal{Physics Letters B}
\makeatletter
\def\ps@pprintTitle{%
  \let\@oddhead\@empty
  \let\@evenhead\@empty
  \let\@oddfoot\@empty
  \let\@evenfoot\@oddfoot
}
\makeatother
\begin{document}
\begin{frontmatter}
\title{Quark Wigner distribution in frame-independent 3-dimensional space}
\author[first]{Sujit Jana}
\author[first]{Vikash Kumar Ojha\corref{cor1}}
\cortext[cor1]{Corresponding author}
\ead{vko@phy.svnit.ac.in}
\affiliation[first]{organization={Department of Physics},%Department and Organization
            addressline={Sardar Vallabhbhai National Institute of Technology}, 
            city={Surat},
            postcode={395 007}, 
            state={Gujarat},
            country={India}}

%%%%%%%%%%%%%%%%%------------Section---------%%%%%%%%%%%%%%%            
\begin{abstract}
We investigate the quark Wigner distribution in a frame-independent, three-dimensional position space within the framework of the dressed quark model. Our findings reveal that the distributions are concentrated near the center of the target and gradually diminish as one moves away in both the longitudinal and transverse directions. The distribution exhibits symmetry along both axes, indicating an equal probability of locating the quark in either direction around the center. Interestingly, the spatial profile of the distribution resembles that of atomic orbitals, where the probability of finding an electron is highest in certain regions compared to others.
%We investigate the spatial distribution of quarks inside inside a target using the dressed quark model. The analysis reveals that the quark distribution is primarily localized around \( b_\perp = 0 \) and \( \sigma = 0 \), with a rapid decline as \( b_\perp \) and \( \sigma \) increase. The distribution exhibits symmetry in both longitudinal (\( \sigma \)) and transverse (\( b_\perp \)) spaces, indicating an equal probability of finding a quark at a given distance from the origin in either direction.
\begin{comment}

    We investigate the three-dimensional Wigner distribution as a function of the boost-invariant variable $\sigma$ and the transverse impact parameter $b_\perp$, for unpolarized, longitudinally polarized, and transversely polarized target state. The boost-invariant coordinate $\sigma=\frac{1}{2}b^-P^+$ is recognized as the Fourier conjugate of the skewness, while the impact parameter $b_\perp$ serves as the Fourier conjugate to the variable $D_\perp=\frac{\Delta_\perp}{1-\xi^2}$, which simplifies to $\Delta_\perp$ when skewness is zero($\xi=0$). Our target is a quark dressed by a gluon, making our system a two-particle system, as we discussed in our previous publication. In this study, we integrated over the quark's transverse momentum $k_\perp$. Thus, the Wigner distributions that we calculate describe the probability density of quarks in hadrons in a three-dimensional light-front coordinate system, offering a unique perspective on the internal structure of hadron.
    \end{comment}
\end{abstract}       
\begin{keyword}
Wigner distribution \sep skewness     
\end{keyword}
\end{frontmatter}

%%%%%%%%%%%%%-------section----------%%%%%%%%%%%%

\section{Introduction}\label{Introduction}
The Wigner distribution offers a semi-classical framework for studying quantum systems by simultaneously incorporating position and momentum space information. For a system described by a wavefunction \( \psi \), the Wigner distribution is defined as  
\[
W(x,p) = \frac{1}{\pi \hbar}\int dy\, \psi^*\left(x - \frac{y}{2}\right) \psi\left(x + \frac{y}{2}\right) e^{i p \cdot y/\hbar},
\]  
which captures the quantum correlations between position and momentum. Originally introduced in quantum mechanics, Wigner distributions have found broad applications across various domains such as quantum optics \cite{optics,optics1,Radhakrishnan:2022khp}, quantum computing \cite{qip,qip1,qip2}, signal processing \cite{sp,sp1,sp2} and quantum chromodynamics \cite{lorce2011quark, mukherjee2014quark, mukherjee2015wigner}. Several methods have been proposed for the direct measurement and experimental reconstruction of the Wigner distribution, particularly in the field of quantum optics \cite{expt,expt1,expt2}.

%Several recent studies have explored their properties and potential experimental realization~\cite{ref1,ref2,ref3,ref4}, with some proposals suggesting that they may, under specific conditions, be measurable in physical systems.

In the context of quantum chromodynamics (QCD), Wigner distributions were first introduced by Ji \cite{Ji2003ak} to investigate the internal structure of hadrons. They are defined through the matrix elements of quark-quark correlators and are connected to generalized transverse momentum-dependent distributions (GTMDs)\cite{Meissner:2008ay, Lorce:2013pza} via Fourier transforms. GTMDs, in turn, relate to generalized parton distributions (GPDs)\cite{Diehl:2003ny,Belitsky:2005qn} and transverse momentum-dependent distributions (TMDs)\cite{Bacchetta:2006tn,Barone:2001sp}, which further reduce to parton distribution functions (PDFs)\cite{Gluck:1998xa, Martin:1998sq} in specific limits. While PDFs offer experimentally extractable and universal one-dimensional information about hadrons, they lack the multidimensional insight necessary for a full spatial and momentum-space characterization.
Wigner distributions, by contrast, provide a richer, multidimensional description of hadronic structure. In this work, we investigate the quark Wigner distribution in a frame-independent, three-dimensional position space within the dressed quark model.

In article \cite{brodsky2006hadron}, the authors present a novel approach to deeply virtual Compton scattering (DVCS) by highlighting an interesting analogy between optical diffraction and features of hadron spectroscopy. The study focuses on DVCS with non-zero longitudinal momentum transfer to the target, offering a more realistic description of the process, as the skewness parameter \( \xi \), which characterizes the longitudinal momentum transfer, is non-zero in any physical experiment. The DVCS amplitude is examined both as a function of \( \xi \) and in the conjugate coordinate space.

To define this coordinate space, consider a position variable \( b \) conjugate to the total momentum transfer \( \Delta \), such that the scalar product is given by
\[
b \cdot \Delta = \frac{1}{2} b^+ \Delta^- + \frac{1}{2} b^- \Delta^+ + \vec{b}_\perp \cdot \vec{\Delta}_\perp.
\]
Since \( \Delta^+ = \xi P^+ \), we can write \( \frac{1}{2} b^- \Delta^+ = \frac{1}{2} b^- P^+ \xi = \sigma \xi \), where \( \sigma \) is interpreted as the longitudinal impact parameter in boost-invariant light-front coordinates. The transverse impact parameters \( \vec{b}_\perp \), conjugate to \( \vec{\Delta}_\perp \), complete the spatial picture. Thus, the distribution in the \( (\vec{b}_\perp, \sigma) \) space yields a unique, frame-independent three-dimensional image of the target.

In \cite{brodsky2007hadron}, the authors further extend this formalism by computing light-front wavefunctions (LFWFs) for hadrons directly in this invariant coordinate space. Motivated by these studies, the present work investigates the quark Wigner distributions within the same boost-invariant three-dimensional position space.

\section{Kinematics and Dressed Quark Model}
We use the light-front coordinate system $(x^+,x^-,x_\perp)$, defining the light-front time and longitudinal spatial coordinates as $x^\pm=x^0\pm x^3$. Additional conventions of light-front coordinates can be found in \cite{harindranath1996introduction,zhang1994light}. 
Our system consists of a quark at one-loop level (serving as the target state) being probed by a virtual photon, which transfers energy \( t = \Delta^2 \) to the target. We denote the initial and final momenta of the target as \( p \) and \( p' \),
\begin{align}\label{kinematics1}
    p=&\Big((1+\xi)P^+,\frac{\Delta_\perp}{2},\frac{m^2+\frac{{\Delta_\perp}^2}{4}}{(1+\xi)P^+}\Big),
    \end{align}
    \begin{align}\label{kinematics2}
    p'=&\Big((1-\xi)P^+,-\frac{\Delta_\perp}{2},\frac{m^2+\frac{{\Delta_\perp}^2}{4}}{(1-\xi)P^+}\Big),
\end{align}
such that the momentum transfer is given by \( \Delta = p - p' \). The parameter \( \xi \), known as the skewness, characterizes the amount of longitudinal momentum transferred to the target. The average longitudinal momentum of the quark is \( k^+ = x P^+ \), where \( x \) is the longitudinal momentum fraction and $P=\frac{p+p'}{2}$ is the target’s average momentum. Since we consider a dressed quark state as the target, the state can be expanded in Fock space up to leading order using light-front wavefunctions
\begin{align}
 \Big{| }p^+,p_\perp,\sigma  \Big{\rangle} = \Phi^\sigma(p) b^\dagger_\sigma(p)
 | 0 \rangle +
 \sum_{\sigma_1 \sigma_2} \int [dp_1]
 \int [dp_2]  \sqrt{16 \pi^3 p^{+}}\nn \\
 \delta^3(p-p_1-p_2)  \Phi^\sigma_{\sigma_1 \sigma_2}(p;p_1,p_2) 
b^\dagger_{\sigma_1}(p_1) 
 a^\dagger_{\sigma_2}(p_2)  | 0 \rangle
 \end{align}
where $[dp]=\frac{dp^+d^2p_\perp}{\sqrt{16\pi^3}p^+}$, and the functions $\Phi^\sigma$, $\Phi^\sigma_{\sigma_1\sigma_2}$ are the light-front wave functions (LFWFs) for a single particle and two particles state. The non-trivial contribution comes from the two-particle LFWF. Using the Jacobi momenta \begin{align}
p^+_i=x_ip^+,\;\;\;\;q_{i\perp}=k_{i\perp}+x_ip_\perp,
\end{align} 
the boost-invariant two-particle LFWF  $\Psi^\sigma_{\sigma_1\sigma_2}(x,q_\perp)=\Phi^\sigma_{\sigma_1,\sigma_2}\sqrt{P^+}$ 
reads \cite{kundu}
\begin{align}
    \Psi^{\sigma a}_{\sigma_1 \sigma_2}(x_,q_{\perp }) = 
\frac{1}{\Big[    m^2 - \frac{m^2 + (q_{\perp })^2 }{x} - \frac{(q_{\perp})^2}{1-x} \Big]}
\frac{g}{\sqrt{2(2\pi)^3}} T^a \chi^{\dagger}_{\sigma_1} \frac{1}{\sqrt{1-x}}
\nn \\ \Big[ 
-2\frac{q_{\perp}}{1-x}   -  \frac{(\sigma_{\perp}.q_{\perp})\sigma_{\perp}}{x}
+\frac{im\sigma_{\perp}(1-x)}{x}\Big]
\chi_\sigma (\epsilon_{\perp \sigma_2})^{*}.
\end{align}
The symbols $\sigma_1, \sigma_2, x, m, \epsilon_{\perp\sigma_2}$ represent the quark's helicity, gluon's helicity, a fraction of the target state's longitudinal momentum, quark's mass, and gluon's polarization vector, respectively.

\begin{comment}
Our system consists of a quark dressed by gluons, interacting with a virtual photon that transfers energy $t=\Delta^2$. The longitudinal momentum transfer is given by $\xi=\frac{\Delta^+}{2P^+}$. For the kinematics, the initial momentum of the dressed quark is denoted as $p$, and the final momentum is represented as $p'$, as shown below
\begin{align}\label{kinematics1}
    p=&\Big((1+\xi)P^+,\frac{\Delta_\perp}{2},\frac{m^2+\frac{{\Delta_\perp}^2}{4}}{(1+\xi)P^+}\Big),
    \end{align}
    \begin{align}\label{kinematics2}
    p'=&\Big((1-\xi)P^+,-\frac{\Delta_\perp}{2},\frac{m^2+\frac{{\Delta_\perp}^2}{4}}{(1-\xi)P^+}\Big).
\end{align}
Here, the target’s average momentum is written as $P=\frac{p+p'}{2}$, and the momentum transfer is defined as $\Delta=p-p'$. The target mass is represented by $m$. The quark’s average four-momentum is symbolized as $k$, where $k^+=xP^+$\cite{ojha2023quark}.
\end{comment}

%%%%%%%%%%%%%-----------section---------%%%%%%%%%%%

\section{GTMDs in Dressed Quark Model}
The quark-quark correlator $W_{\lambda,\lambda'}^\Gamma(x,\xi,\Delta_\perp,k_\perp;S)$ is defined through the non-diagonal matrix element of the bi-local quark field \cite{mukherjee2014quark,meissner2009generalized} 
\begin{align}\label{qqc}
    W_{\lambda,\lambda'}^{[\Gamma]}(x,\xi,\Delta_\perp,k_\perp;S)=&\frac{1}{2}\int\frac{dz^-}{2\pi}\frac{d^2z_\perp}{(2\pi)^2}e^{ip.z}\Big<p',\lambda'\Big|\Bar{\psi}(-\frac{z}{2})\nn\\
    &\mathcal{W}_{[-\frac{z}{2},\frac{z}{2}]}\Gamma\psi(\frac{z}{2})\Big|p,\lambda\Big>\Bigg|_{z^+=0}.
\end{align}
Here $\Delta_\perp$ is the total transverse momentum transferred, and $\xi$ (skewness) is the fraction of longitudinal momentum transferred to the target. The state $|p,\lambda\rangle$ refers to the initial, and $|p',\lambda'\rangle$ represents the final state of the target, where $\lambda$ and $\lambda'$ indicate their respective helicities. The Wilson line 
$\mathcal{W}_{[-\frac{z}{2},\frac{z}{2}]}$ serves as a gauge link between the two quark fields 
$\psi(\frac{z}{2})$ and $\Bar{\psi}(-\frac{z}{2})$, while 
$\Gamma$ belongs to the set 
$\{\gamma^+,\gamma^+\gamma^5,i\sigma^{+j}\gamma^5\}$, corresponding to the unpolarized, longitudinally polarized, and transversely polarized quark. The quark-quark correlator in Eq. (\ref{qqc}) can be parameterized in terms of generalized transverse momentum dependent parton distributions (GTMDs) for unpolarized, longitudinally polarized, and transversely polarized dressed quarks as follows \cite{meissner2009generalized}:
\begin{align}\label{unpara}
W^{[\gamma^+]}_{\lambda,\lambda'}=&\frac{1}{2m}\Bar{u}(p',\lambda')\Big[F_{1,1}-\frac{i\sigma^{i+}k_{i\perp}}{P^+}F_{1,2}-\frac{i\sigma^{i+}\Delta_{i\perp}}{P^+}F_{1,3}\nn\\
&+\frac{i\sigma^{ij}k_{i\perp}\Delta_{j\perp}}{m^2}F_{1,4}\Big]u(p,\lambda),\\
W^{[\gamma^{+}\gamma_5]}_{\lambda,\lambda'}=&\frac{1}{2m}\Bar{u}(p',\lambda')\Big[\frac{-i\epsilon^{ij}_{\perp}k_{i\perp}\Delta_{j\perp}}{m^2}G_{1,1}-\frac{i\sigma^{i+}\gamma_5 k_{i\perp}}{P^+}G_{1,2}\nn\\
&-\frac{i\sigma^{i+}\gamma_5 \Delta_{i\perp}}{P^+}G_{1,3}+i\sigma^{+-}\gamma_5 G_{1,4}\Big]u(p,\lambda),\\
W^{[i\sigma^{+j}\gamma_5]}_{\lambda\lambda'}=&\frac{1}{2m}\Bar{u}(p',\lambda')\Big[-\frac{i\epsilon^{ij}_\perp p^i_\perp}{m}H_{1,1}-\frac{i\epsilon^{ij}_\perp \Delta^i_\perp}{m}H_{1,2}\nn\\
&+\frac{mi\sigma^{j+}\gamma^5}{P^+}H_{1,3}+\frac{p^j_\perp i \sigma^{k+}\gamma^5p^k_\perp}{mP^+}H_{1,4}\nn\\
&+\frac{\Delta^j_\perp i \sigma^{k+}\gamma^5p^k_\perp}{mP^+}H_{1,5}+\frac{\Delta^j_\perp i \sigma^{k+}\gamma^5\Delta^k_\perp}{mP^+}H_{1,6}\nn\\
&+\frac{p^j_\perp i \sigma^{+-}\gamma^5}{m}H_{1,7}+\frac{\Delta^j_\perp i \sigma^{+-}\gamma^5}{m}H_{1,8}\Big]u(p,\lambda),\label{tpara}
\end{align}
the functions $F_{1,i}$, $G_{1,i}$, $H_{1,j}$, where $i=1,2,...,4$ and $j=1,2,...,8$ are the GTMDs for quark. These GTMDs reduced to TMDs and GPDs under some integral limit and have been studied in the different model\cite{maji2022leading,mukherjee2014quark}. The expression for GTMDs in dressed quark model can be obtained using the Fock state expansion of target state and the Light-front wavefunctions. The analytical expression for GTMDs for zero and non-zero skewness in the dressed quark model are presented in \cite{mukherjee2014quark, ojha2023quark}. To get the Wigner distribution for quark in frame-independent 3-dimensional position space, we use these results of GTMDs for non-zero skewness in the dressed quark model.  

%%%%%%%%%%%%---------section-----------%%%%%%%%

\section{Wigner distribution and GTMDs for non-zero skewness}\label{WDGNS}

%----------- Trial
The Wigner distribution of quarks for non-zero skewness can be defined as the two-dimensional Fourier transform of the generalized transverse momentum distributions (GTMDs) \cite{lorce2011quark,meissner2009generalized}.

%------------------
%\begin{comment}
%The Wigner distribution of quarks can be defined as a function of the transverse impact parameter ($b_\perp$), obtained through a two-dimensional Fourier transform of the generalized transverse momentum distributions (GTMDs) \cite{lorce2011quark,meissner2009generalized}

\begin{align}
\rho^{[\Gamma]}(x,\xi,b_\perp,k_\perp;S)&=\int\frac{d^2D_\perp}{(2\pi)^2}e^{iD_\perp\cdot b_\perp}W_{\lambda,\lambda'}^{[\Gamma]}(x,\xi,\Delta_\perp,k_\perp;S)\label{rho_b}.
\end{align}
where the transverse impact parameter $b_\perp$ is the Fourier conjugate of the variable $D_\perp=\frac{\Delta_\perp}{1-\xi^2}$, which becomes $\Delta_\perp$ when the skewness is zero $(\xi=0)$. The quark-quark correlator $W_{\lambda,\lambda'}^{[\Gamma]}(x,\xi,\Delta_\perp,k_\perp;S)$ is related to the GTMDs through equation (7-9).  Here the Fourier transform of the correlator function $W_{\lambda,\lambda'}^{[\Gamma]}(x,\xi,\Delta_\perp,k_\perp;S)$ with respect to $D_\perp$ gives a distribution in transverse impact parameter space $b_\perp$. Similarly, we can define the Wigner distribution for a quark in longitudinal impact parameter space as \cite{maji2022leading,jana2025gluon}

\begin{align} \label{rho_sigma}
 \rho^{[\Gamma]}(x,\sigma,\Delta_\perp,k_\perp;S)&=\int\frac{d\xi}{2\pi}e^{i\sigma\cdot\xi}W_{\lambda,\lambda'}^{[\Gamma]}(x,\xi,\Delta_\perp,k_\perp;S).
\end{align}
where the skewness variable $(\xi)$ is Fourier conjugate to the boost-invariant longitudinal impact parameter, which is defined as $\sigma=\frac{1}{2}b^-P^+$. Here, the Fourier transformation of the correlator function $W^{[\Gamma]}(x,\xi,\Delta_\perp,k_\perp;S)$ with respect to the skewness variable $\xi$ reveals a distribution in boost-invariant longitudinal impact parameter space $\sigma$.
%\end{comment}
\begin{figure}[htp!]
\begin{minipage}[c]{1\textwidth}
\small{(a)}\includegraphics[width=8cm,height=5cm,clip]{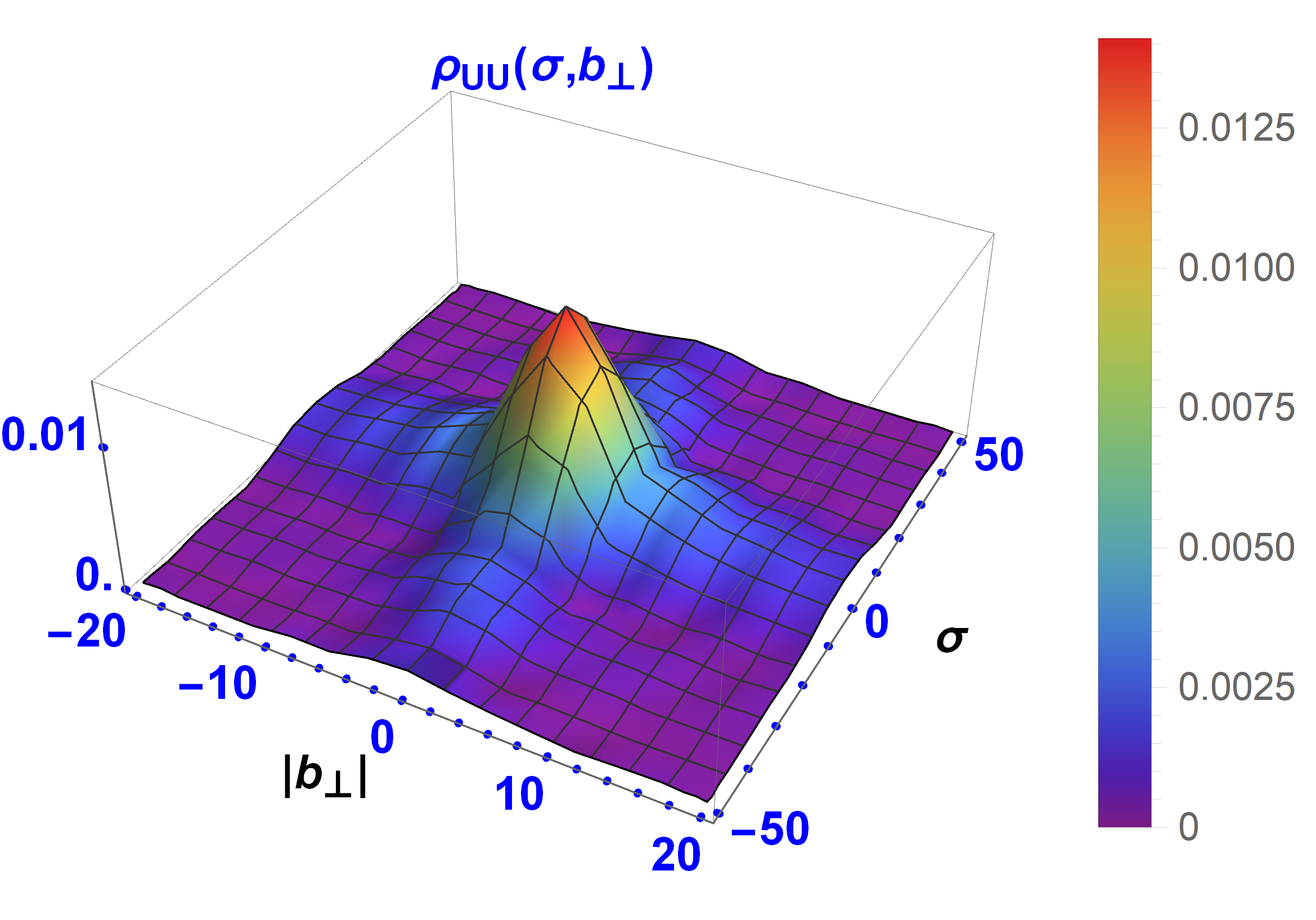}\\
\small{(b)}\includegraphics[width=8cm,height=5cm,clip]{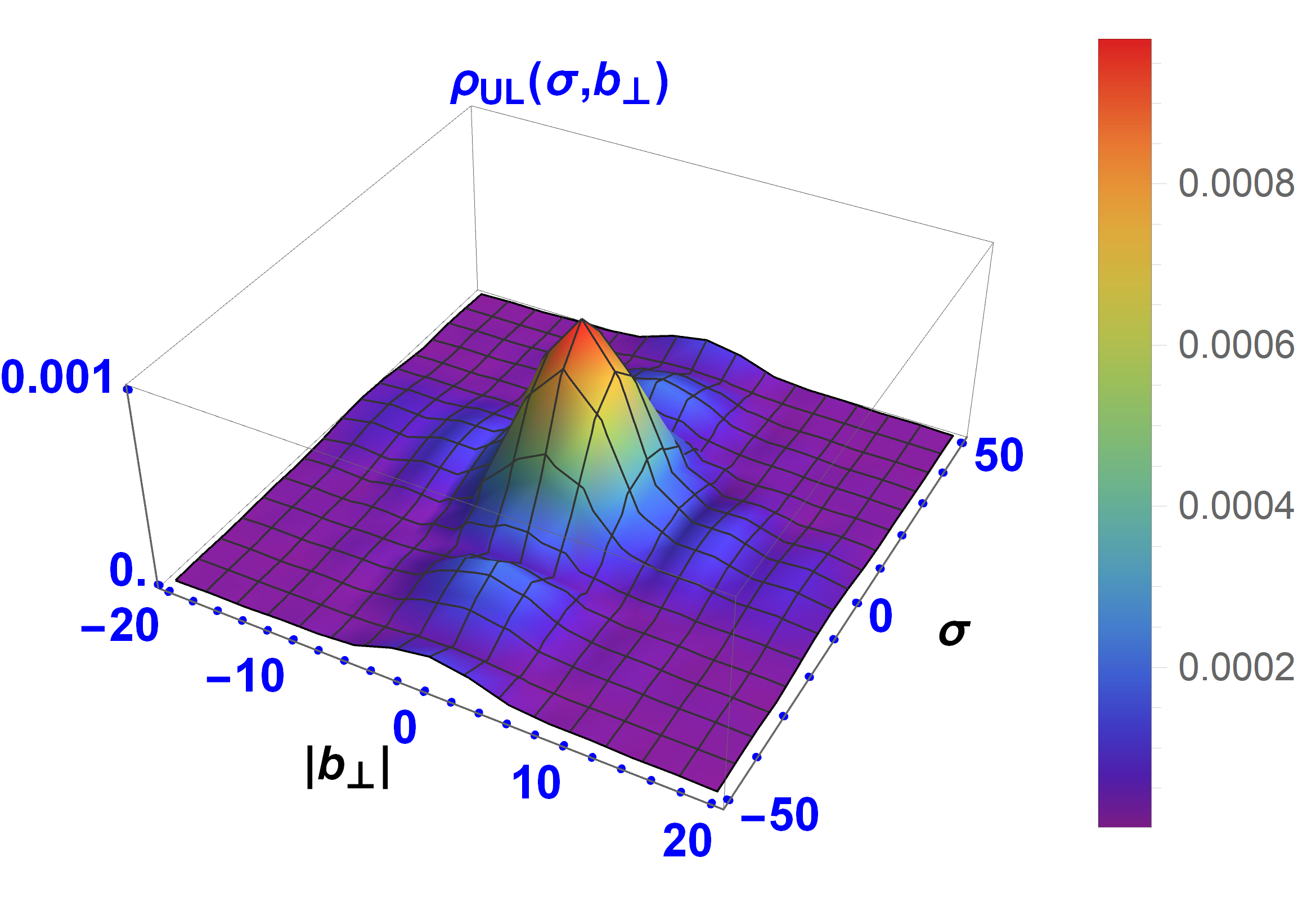}\\
\small{(c)}\includegraphics[width=8cm,height=5cm,clip]{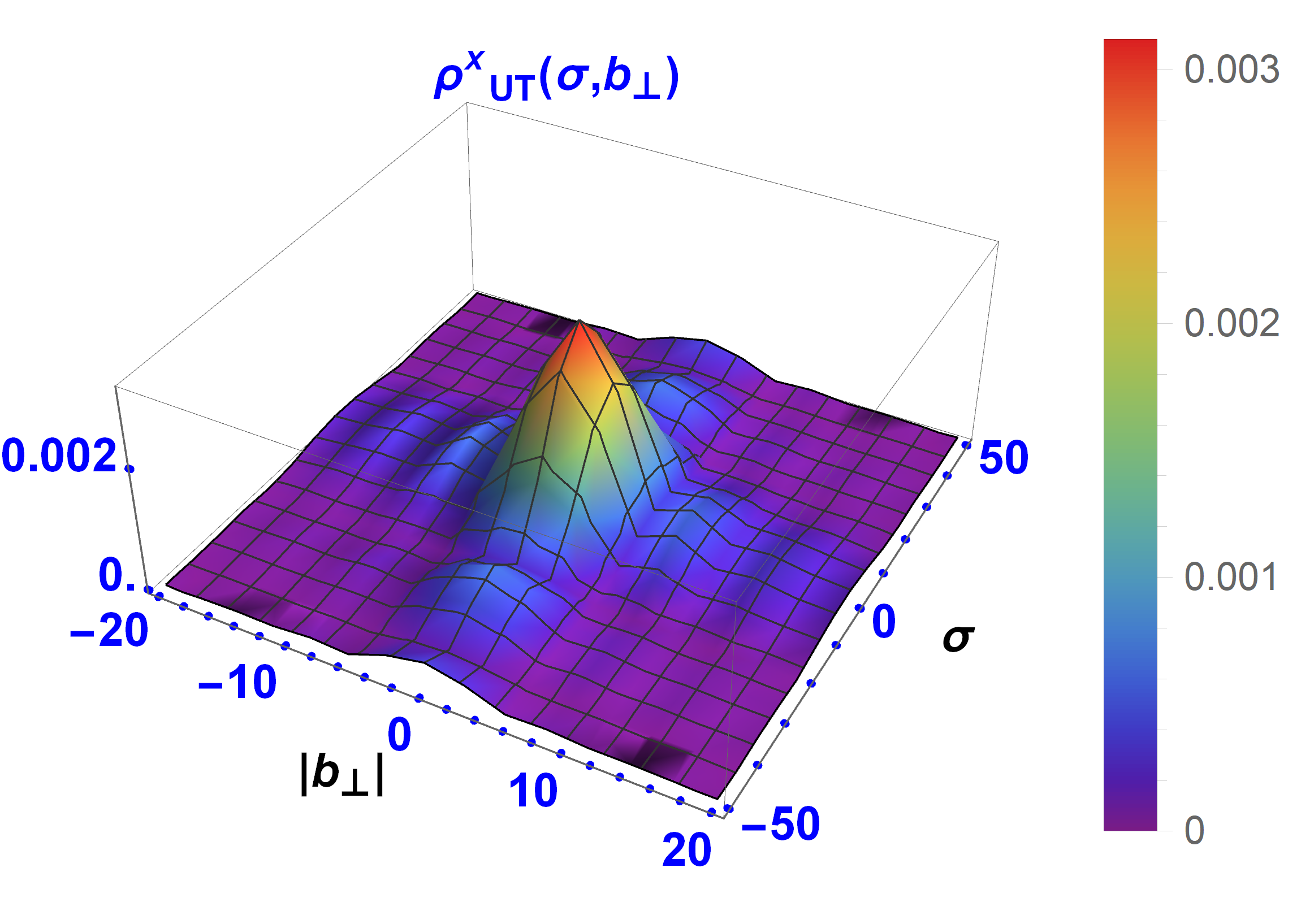}\\
\small{(d)}\includegraphics[width=8cm,height=5cm,clip]{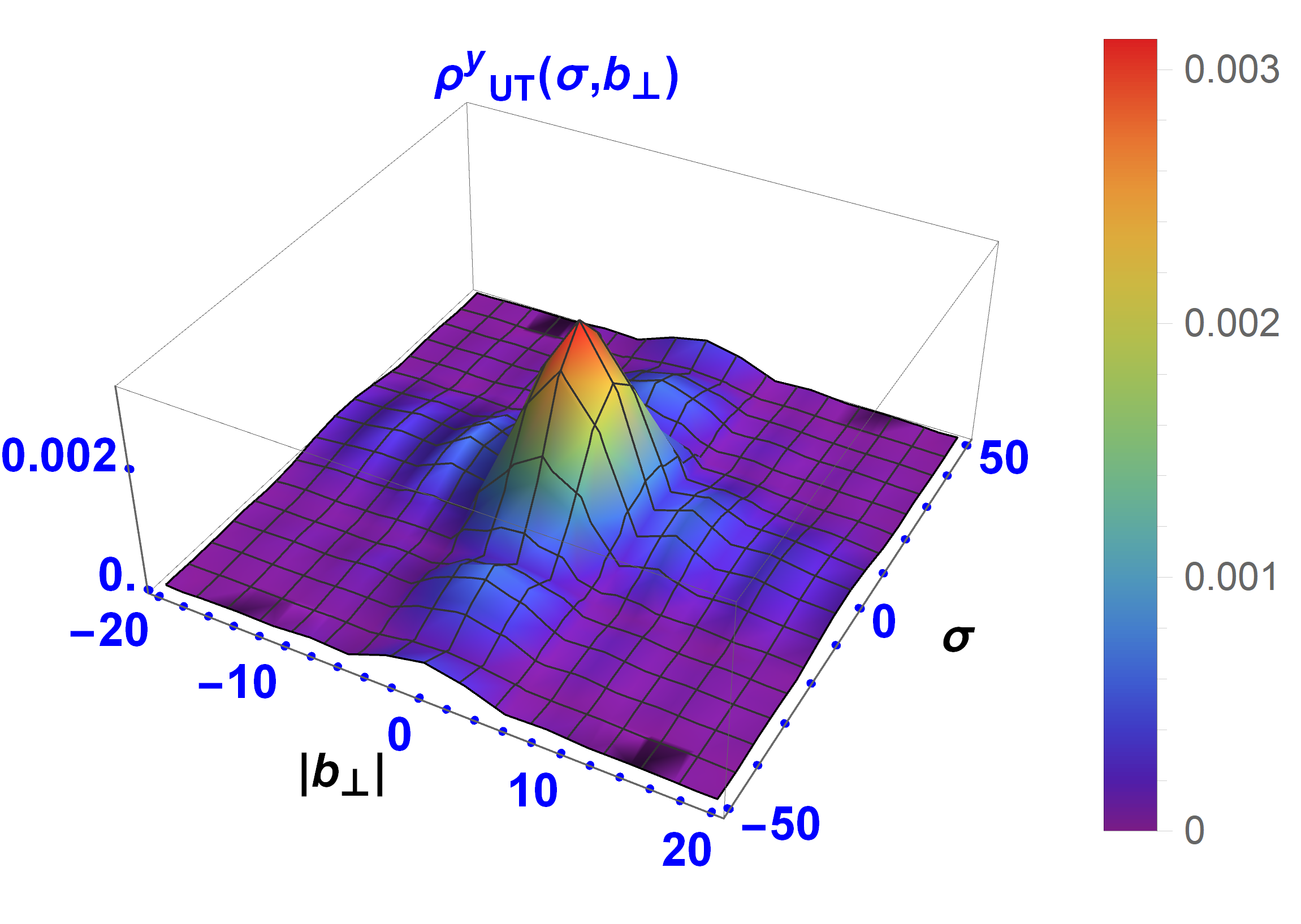}
\end{minipage}
\caption{\label{FigWigun}Quark Wigner distributions in the unpolarized target for various quark polarization states: (a) unpolarized, (b) longitudinally polarized, and (c), (d) transversely polarized.}
\end{figure}
%%%%%%%%%%%%%%----------section---------%%%%%%%%%%%

\section{Quark Wigner distribution in 3-D space}\label{QD3S}
Combining Eq.(\ref{rho_b}) and Eq.(\ref{rho_sigma}), we can define the Wigner distribution as function of both longitudinal and transverse impact parameters,
\begin{align}
\rho^{[\Gamma]}(x,\sigma,b_\perp;S)=&\int d^2k_\perp\int\frac{d\xi}{2\pi}e^{i\sigma\cdot\xi}\int\frac{d^2\Delta_\perp}{1-\xi^2}e^{i\frac{b_\perp\cdot\Delta_\perp}{1-\xi^2}}\nn\\
&W_{\lambda,\lambda'}^{[\Gamma]}(x,\xi,\Delta_\perp,k_\perp;S)\nn\\
=&\int d^2k_\perp\int\frac{d\xi}{2\pi}\int\frac{d^2\Delta_\perp}{(1-\xi^2)}e^{i(\sigma.\xi+\frac{b_\perp.\Delta_\perp}{1-\xi^2})}\nn\\
&W_{\lambda,\lambda'}^{[\Gamma]}(x,\xi,\Delta_\perp,k_\perp;S).\label{QWD3D}
\end{align}
Here, $S$ represents the polarization of the dressed quark system state. Additional Wigner distributions \( \rho_{XY} \) can be defined based on the polarization states of both the target and the struck quark, where \( X \) and \( Y \) represent the polarization of the dressed quark (target) and the struck quark, respectively. For unpolarized, longitudinally polarized, and transversely polarized targets, the corresponding Wigner distributions are

%Additional Wigner distributions can be defined based on the polarization states of both the target and the struck quark 
\begin{align}
\rho_{UY}(x,\sigma,b_\perp)=&\frac{1}{2}\Big[\rho^{[\Gamma]}(x,\sigma,b_\perp,+\hat{e}_z)+\rho^{[\Gamma]}(x,\sigma,b_\perp,-\hat{e}_z)\Big]\label{un_1},\\
\rho_{LY}(x,\sigma,b_\perp)=&\frac{1}{2}\Big[\rho^{[\Gamma]}(x,\sigma,b_\perp,+\hat{e}_z)-\rho^{[\Gamma]}(x,\sigma,b_\perp,-\hat{e}_z)\Big]  \label{long_1}, \\
\rho^i_{TY}(x,\sigma,b_\perp)=&\frac{1}{2}\Big[\rho^{[\Gamma]}(x,\sigma,b_\perp,+\hat{e}_i)-\rho^{[\Gamma]}(x,\sigma,b_\perp,-\hat{e}_i)\Big]    \label{trans_1}.
\end{align}
The operator \( \Gamma \) is chosen from the set \( \{ \gamma^+,\, \gamma^+\gamma^5,\, i\sigma^{+j}\gamma^5 \} \), depending on the polarization state of the struck quark. In Eq.(\ref{trans_1}), the index $i=\hat{x},\;\hat{y}$ specifies the direction of the target state's transverse polarization within the transverse plane.
%
%where $X$ and $Y$ represent the polarization of the dressed quark (target) and the struck quark, respectively. The subscripts $X,\;Y=U,\;L,\;T$ correspond to unpolarized, longitudinally polarized, and transversely polarized cases. 
%
All Wigner distributions defined in Eqs. (\ref{un_1}-\ref{trans_1}) can be expressed in terms of GTMDs. For instance, $\rho_{UU}$ can be simplified using Eq.(\ref{QWD3D}) to obtain the following expression 
%
%Each of the equations Eqs.(\ref{un_1}-\ref{trans_1}) combines eight Wigner distributions, corresponding to the three distinct polarizations of the quark $Y={U,\;L,\;T}$ and their associated gamma structures $\Gamma=\{\gamma^+,\; \gamma^+\gamma^5,\;i\sigma^{+j}\gamma^5\}$, respectively. In Eq.(\ref{trans_1}), the index $i=\hat{x},\;\hat{y}$ indicates the direction of the transversely polarized dressed quark system in the transverse plane. Using Eq.(\ref{QWD3D}) and Eq.(\ref{un_1}), we can calculate the Wigner distribution $\rho_{UY}$ for $Y=U$ as shown below
\begin{align}
 \rho_{UU}(x,\sigma,b_\perp)=&\frac{1}{2}\Big[\rho^{[\gamma^+]}(x,\sigma,b_\perp,+\hat{e}_z)+\rho^{[\gamma^+]}(x,\sigma,b_\perp,-\hat{e}_z)\Big]\nn\\
 =&\frac{1}{2}\int d^2k_\perp\int\frac{d\xi}{2\pi}\int\frac{d^2\Delta_\perp}{(1-\xi^2)}e^{i(\sigma.\xi+\frac{b_\perp.\Delta_\perp}{1-\xi^2})}\nn\\
 &\Big[W_{\uparrow\uparrow}^{[\gamma^+]}(x,\xi,\Delta_\perp,k_\perp)+W_{\downarrow\downarrow}^{[\gamma^+]}(x,\xi,\Delta_\perp,k_\perp)\Big]\nn\\
 %=&\frac{1}{2}\int d^2k_\perp\int\frac{d\xi}{2\pi}\int\frac{d^2\Delta_\perp}{(1-\xi^2)}e^{i(\sigma.\xi+\frac{b_\perp.\Delta_\perp}{1-\xi^2})}\nn\\
 %&\frac{2F_{1,1}}{\sqrt{1-\xi^2}}\nn\\
 =&\int d^2k_\perp\int\frac{d\xi}{2\pi}\int\frac{d^2\Delta_\perp}{(2\pi)^2}\frac{1}{(1-\xi^2)^{\frac{3}{2}}}\nn\\
&e^{i(\sigma.\xi+\frac{b_\perp.\Delta_\perp}{1-\xi^2})}F_{1,1}.
\end{align}
In a similar manner, the remaining Wigner distributions can be derived in terms of the corresponding GTMDs as follows
\begin{align}
\rho_{UU}(x,\sigma,b_\perp)=&\int d^2k_\perp\int\frac{d\xi}{2\pi}\int\frac{d^2\Delta_\perp}{(2\pi)^2}\frac{1}{(1-\xi^2)^{\frac{3}{2}}}\nn\\
&e^{i(\sigma.\xi+\frac{b_\perp.\Delta_\perp}{1-\xi^2})}F_{1,1}\label{eqn.rhuu},
\end{align}
\begin{align}
\rho_{UL}(x,\sigma,b_\perp)=&\int d^2k_\perp\int\frac{d\xi}{2\pi}\int\frac{d^2\Delta_\perp}{(2\pi)^2}\frac{-i}{m^2(1-\xi^2)^{\frac{3}{2}}}\nn\\
&\epsilon^{ij}_\perp k^i_\perp\Delta^j_\perp e^{i(\sigma.\xi+\frac{b_\perp.\Delta_\perp}{1-\xi^2})} G_{1,1},\\
\rho^j_{UT}(x,\sigma,b_\perp)=&\int d^2k_\perp\int\frac{d\xi}{2\pi}\int\frac{d^2\Delta_\perp}{(2\pi)^2}\frac{-i}{m^2(1-\xi^2)^{\frac{3}{2}}}\nn\\
&\epsilon^{ij}_\perp e^{i(\sigma.\xi+\frac{b_\perp.\Delta_\perp}{1-\xi^2})}\Big[k^i_\perp H_{1,1}+\Delta^i_\perp H_{1,2}\Big],\\
\rho_{LU}(x,\sigma,b_\perp)=&\int d^2k_\perp\int\frac{d\xi}{2\pi}\int\frac{d^2\Delta_\perp}{(2\pi)^2}\frac{i}{m^2(1-\xi^2)^{\frac{3}{2}}}\nn\\
&\epsilon^{ij}_\perp k^i_\perp\Delta^j_\perp e^{i(\sigma.\xi+\frac{b_\perp.\Delta_\perp}{1-\xi^2})}F_{1,4},\\
\rho_{LL}(x,\sigma,b_\perp)=&\int d^2k_\perp\int\frac{d\xi}{2\pi}\int\frac{d^2\Delta_\perp}{(2\pi)^2}\frac{2}{(1-\xi^2)^{\frac{3}{2}}}\nn\\
&e^{i(\sigma.\xi+\frac{b_\perp.\Delta_\perp}{1-\xi^2})}G_{1,4},\\
\rho^j_{LT}(x,\sigma,b_\perp)=&\int d^2k_\perp\int\frac{d\xi}{2\pi}\int\frac{d^2\Delta_\perp}{(2\pi)^2}\frac{2}{m(1-\xi^2)^{\frac{3}{2}}}\nn\\
&e^{i(\sigma.\xi+\frac{b_\perp.\Delta_\perp}{1-\xi^2})}\Big[k^j_\perp H_{1,7}+\Delta^j_\perp H_{1,8}\Big],\\
\rho^i_{TU}(x,\sigma,b_\perp)=&\int d^2k_\perp\int\frac{d\xi}{2\pi}\int\frac{d^2\Delta_\perp}{(2\pi)^2}\frac{-i}{2m(1-\xi^2)^{\frac{3}{2}}}\nn\\
&\epsilon^{ij}_\perp e^{i(\sigma.\xi+\frac{b_\perp.\Delta_\perp}{1-\xi^2})}\Big[\Delta^j_\perp(F_{1,1}-2(1-\xi^2)F_{1,3})-\nn\\
&2(1-\xi^2)k^j_\perp F_{1,2}+\frac{\xi}{m^2}\epsilon^{k,l}_\perp k^k_\perp\Delta^l_\perp\Delta^j_\perp F_{1,4}\Big],\\
\rho^j_{TL}(x,\sigma,b_\perp)=&\int d^2k_\perp\int\frac{d\xi}{2\pi}\int\frac{d^2\Delta_\perp}{(2\pi)^2(1-\xi^2)}\nn\\
&e^{i(\sigma.\xi+\frac{b_\perp.\Delta_\perp}{1-\xi^2})}\Big[\frac{-1}{2m^3(1-\xi^2)^{\frac{3}{2}}}\epsilon^{ij}_\perp\epsilon^{kl}_\perp k^k_\perp\Delta^l_\perp\Delta^j_\perp\nn\\
& G_{1,1}+\frac{\sqrt{1-\xi^2}}{m}k^i_\perp G_{1,2}+\frac{1}{m\sqrt{1-\xi^2}}\Delta^i_\perp\nn\\
&((1-\xi^2)G_{1,3}-\xi G_{1,4})\Big]\label{eqn.rhTL}.
%\rho^j_{TT}(x,\sigma,b_\perp)=&N\int d^2k_\perp\int\frac{d\xi}{2\pi}\int\frac{d^2\Delta_\perp}{(2\pi)^2(1-\xi^2)}\epsilon^{ij}_\perp(-1)^j\nn\\
%&e^{i(\sigma.\xi+\frac{b_\perp.\Delta_\perp}{1-\xi^2})}\Big[\frac{1}{2m^2\sqrt{1-\xi^2}}( k^i_\perp\Delta^i_\perp H_{1,1}+(\Delta^i_\perp)^2\nn\\
%&H_{1,2})+\sqrt{1-\xi^2}H_{1,3}+\frac{\sqrt{1-\xi^2}}{m^2}(k^j_\perp)^2H_{1,4}\nn\\
%&+\frac{1}{m^2\sqrt{1-\xi^2}}k^j_\perp\Delta^j_\perp((1-\xi^2)H_{1,5}-\xi H_{1,7})\nn\\
%&+\frac{1}{m^2\sqrt{1-\xi^2}}(\Delta^j_\perp)^2((1-\xi^2)H_{1,6}\nn\\
%&-\xi H_{1,8})\Big]\label{eqn.rhTT}.
\end{align}
The GTMDs $F_{1,i}$, $G_{1,i}$, $H_{1,j}$ are defined in Eq. (\ref{unpara}-\ref{tpara}). The analytical expressions of these GTMDs for quarks in the dressed quark model at zero skewness were derived and presented in \cite{mukherjee2014quark}, whereas the corresponding expressions for non-zero skewness were obtained and reported in \cite{ojha2023quark}. In the following section, we use the results from \cite{ojha2023quark}, i.e the analytical expression of quark's GTMDs to derive the quark Wigner distributions in three-dimensional boost-invariant space.

%The analytical expressions of these GTMDs for quarks in the dressed quark model at zero skewness were provided in \cite{mukherjee2014quark}, while the corresponding expressions for non-zero skewness are presented in \cite{ojha2023quark}. 
%Additionally, $\epsilon^{ij}_\perp$ is defined according to the paper \cite{kaur2018wigner}, while $m$ stands for the quark mass, set at 0.0033 GeV.

\begin{comment}   
\begin{figure}[htp!]
\begin{minipage}[c]{1\textwidth}
\small{(a)}\includegraphics[width=8cm,height=5cm,clip]{Plots/rhoUUmixspace5.png}\\
\small{(b)}\includegraphics[width=8cm,height=5cm,clip]{Plots/rhoULmixspace5.png}\\
\small{(c)}\includegraphics[width=8cm,height=5cm,clip]{Plots/rhoUTxmixspace5.png}\\
\small{(d)}\includegraphics[width=8cm,height=5cm,clip]{Plots/rhoUTymixspace5.png}
\end{minipage}
\caption{\label{FigWigun}Quark Wigner distributions in the unpolarized target for various quark polarization states: (a) unpolarized, (b) longitudinally polarized, and (c), (d) transversely polarized.}
\end{figure}
\end{comment}

\begin{figure}[htp!]
\begin{minipage}[c]{1\textwidth}
\small{(a)}\includegraphics[width=8cm,height=5cm,clip]{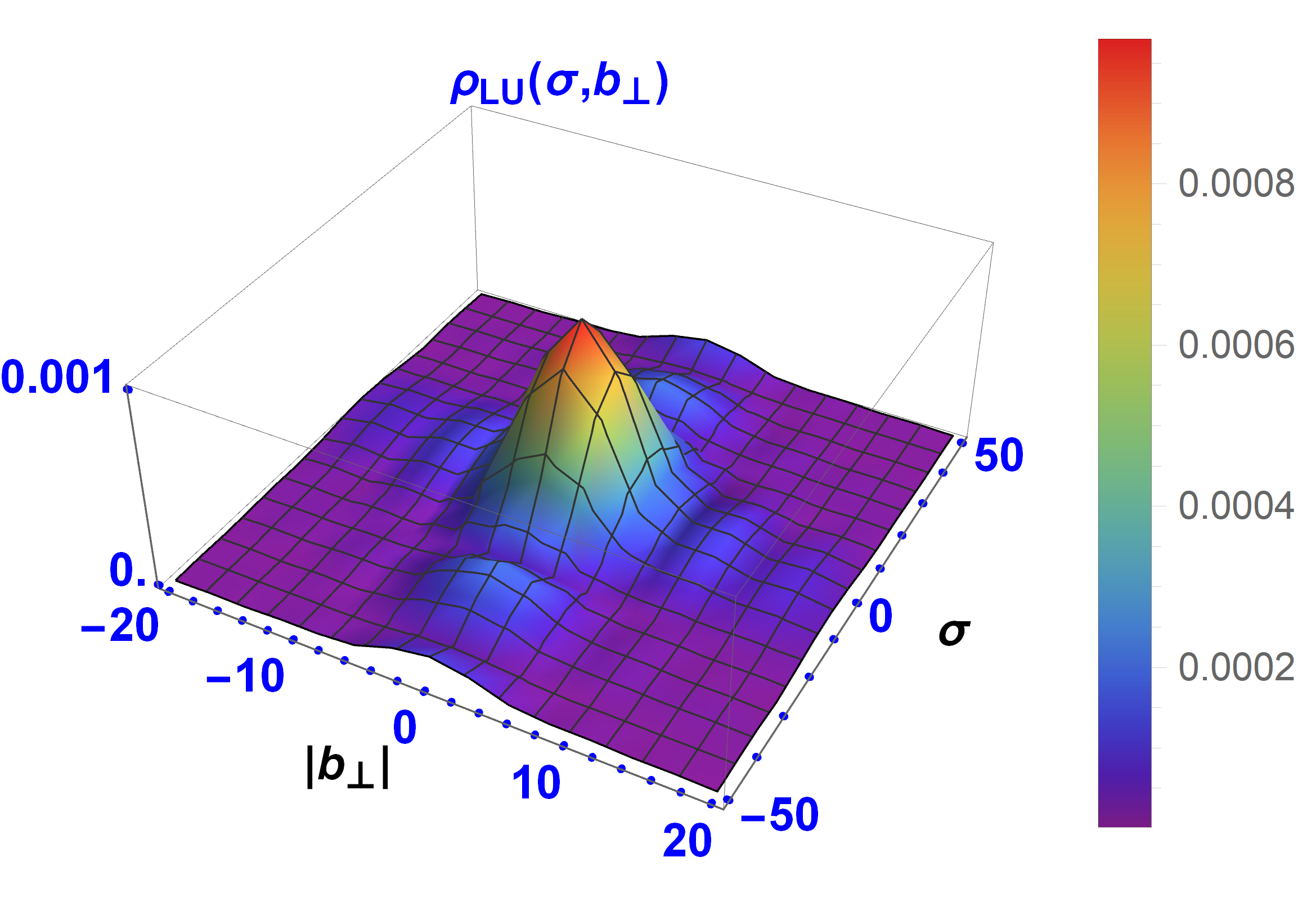}\\
\small{(b)}\includegraphics[width=8cm,height=5cm,clip]{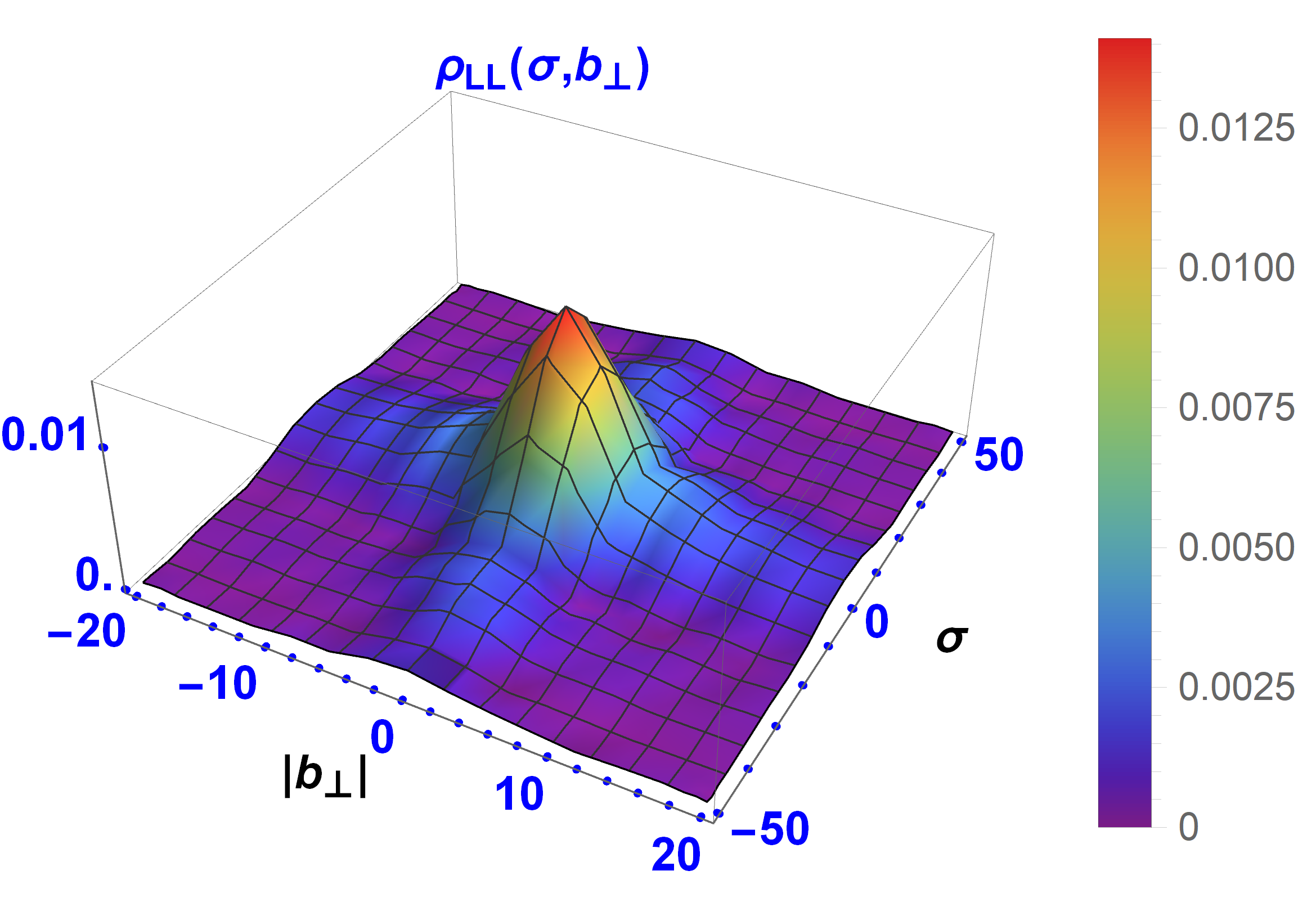}\\
\small{(c)}\includegraphics[width=8cm,height=5cm,clip]{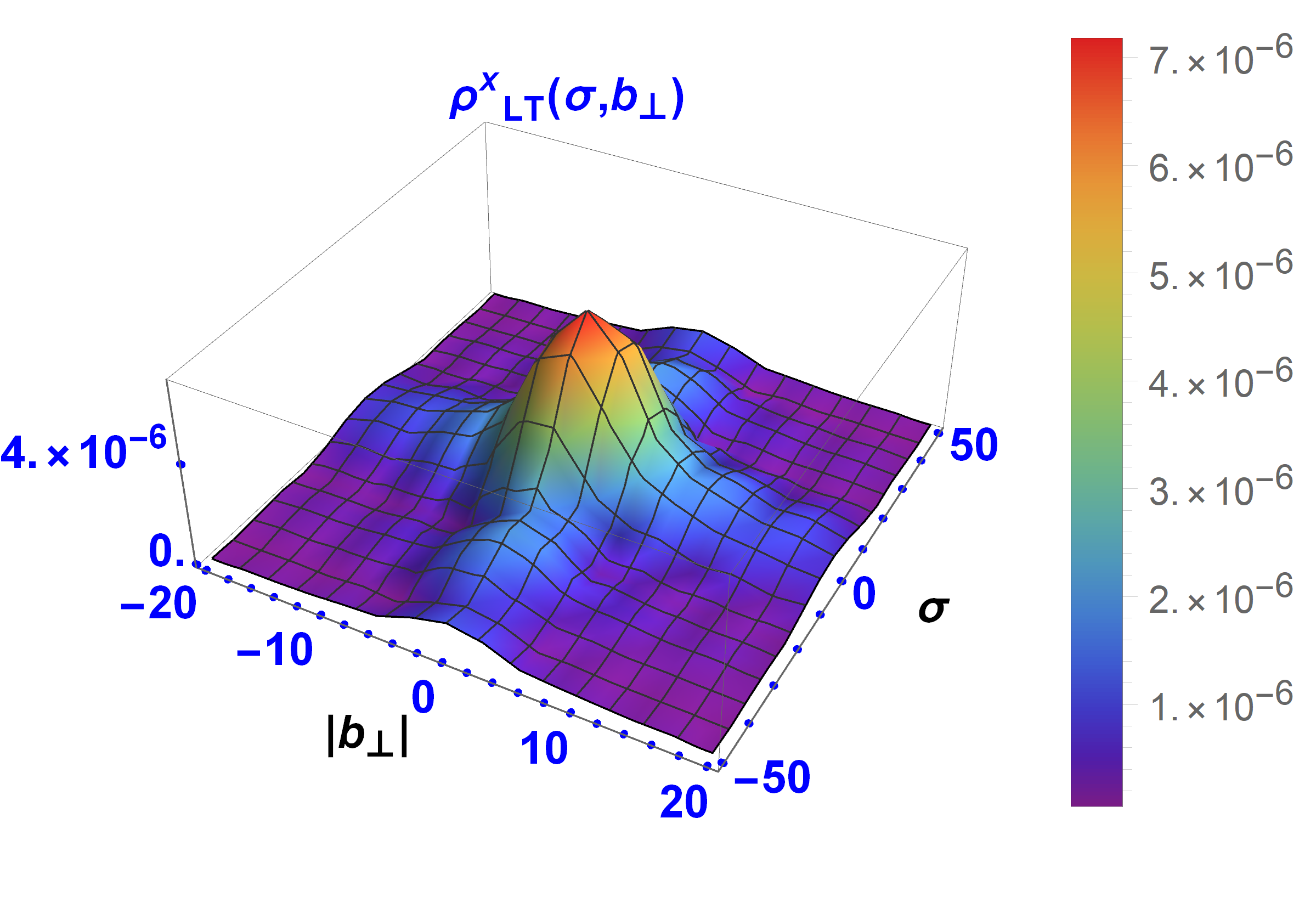}\\
\small{(d)}\includegraphics[width=8cm,height=5cm,clip]{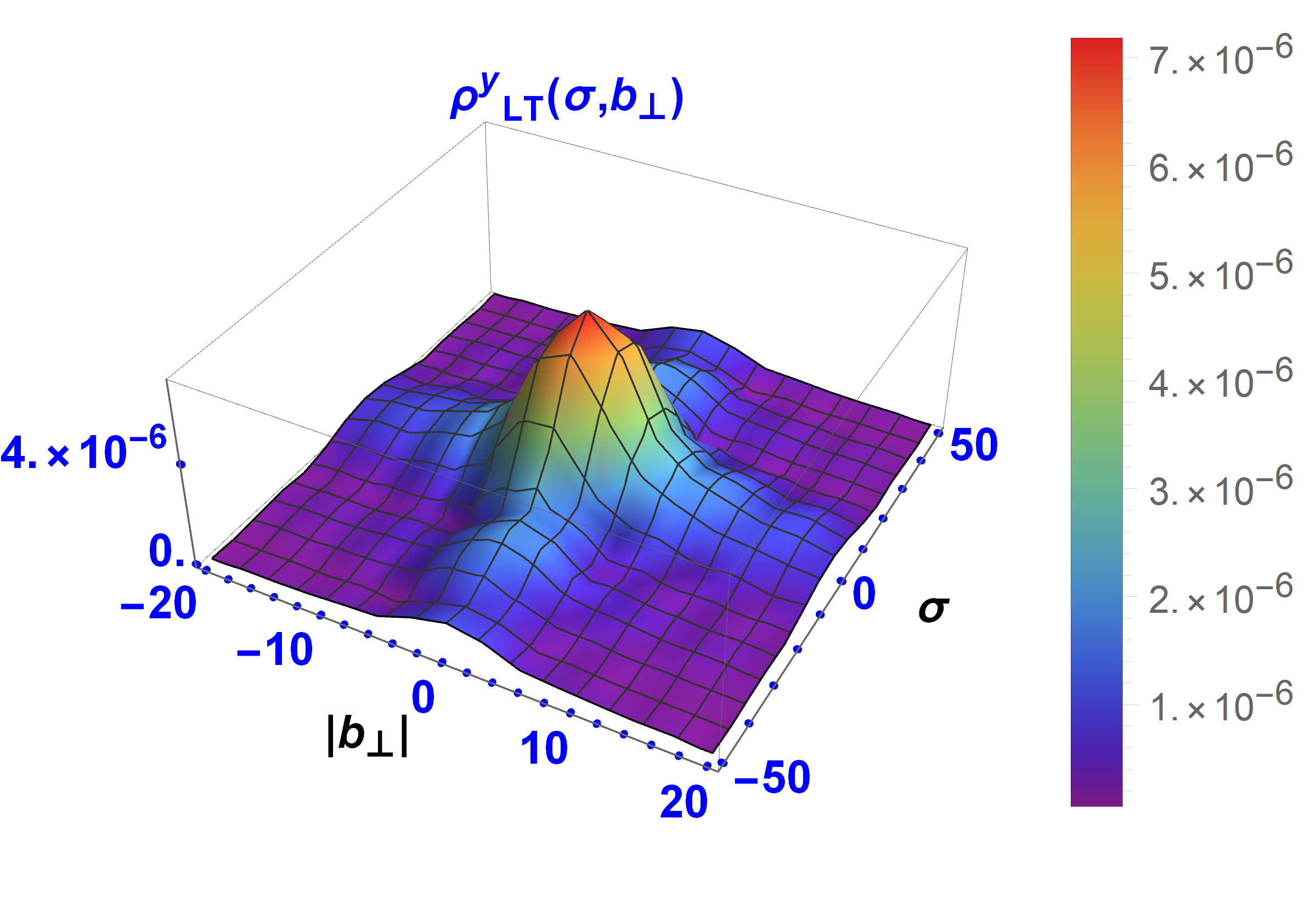}
\end{minipage}
\caption{\label{FigWiglong} Quark Wigner distributions in the longitudinally polarized target for various quark polarization states: (a) unpolarized, (b) longitudinally polarized, and (c), (d) transversely polarized.}
\end{figure}
\begin{figure}[htp!]
\begin{minipage}[c]{1\textwidth}
\small{(a)}\includegraphics[width=8cm,height=5cm,clip]{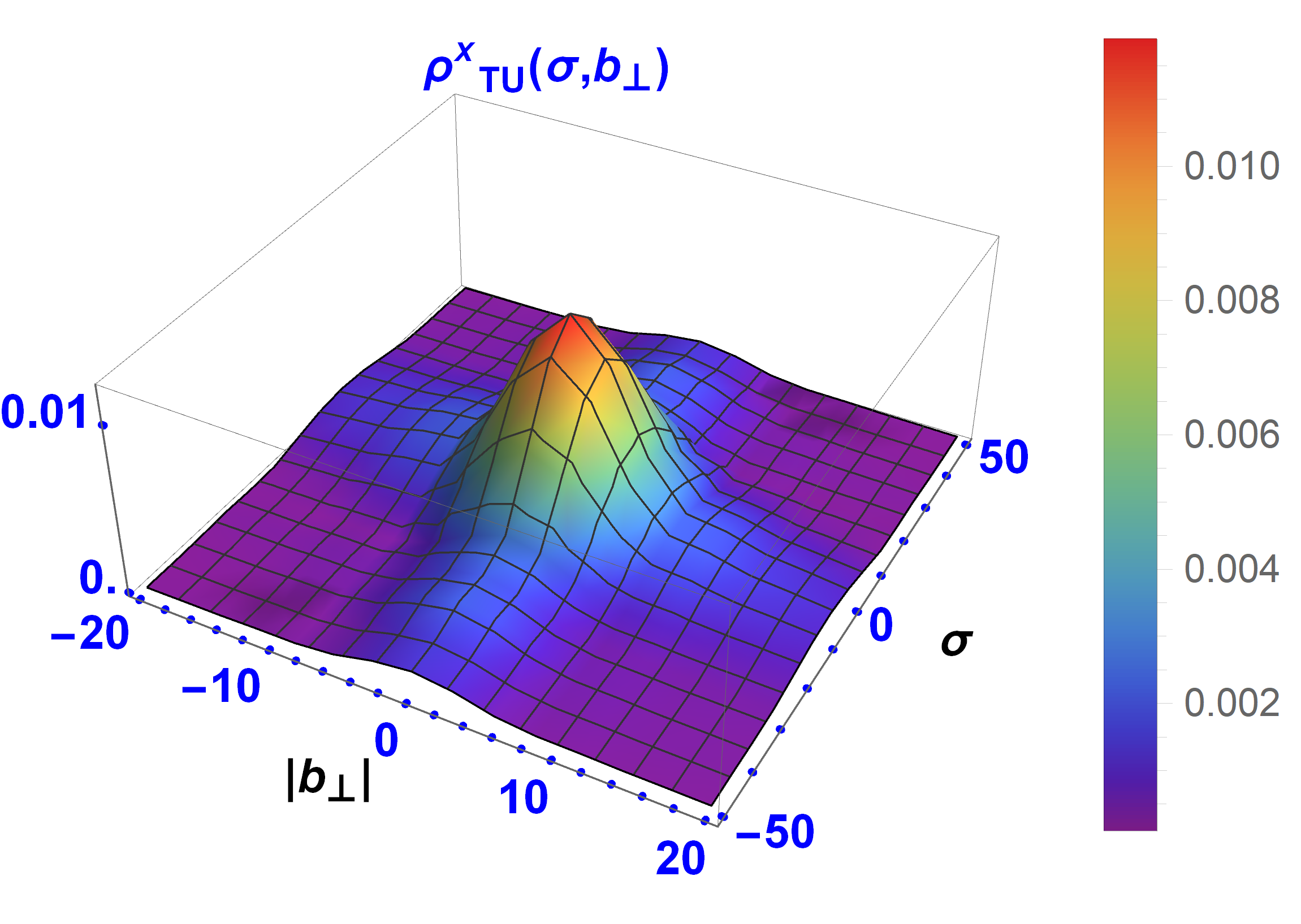}\\
\small{(b)}\includegraphics[width=8cm,height=5cm,clip]{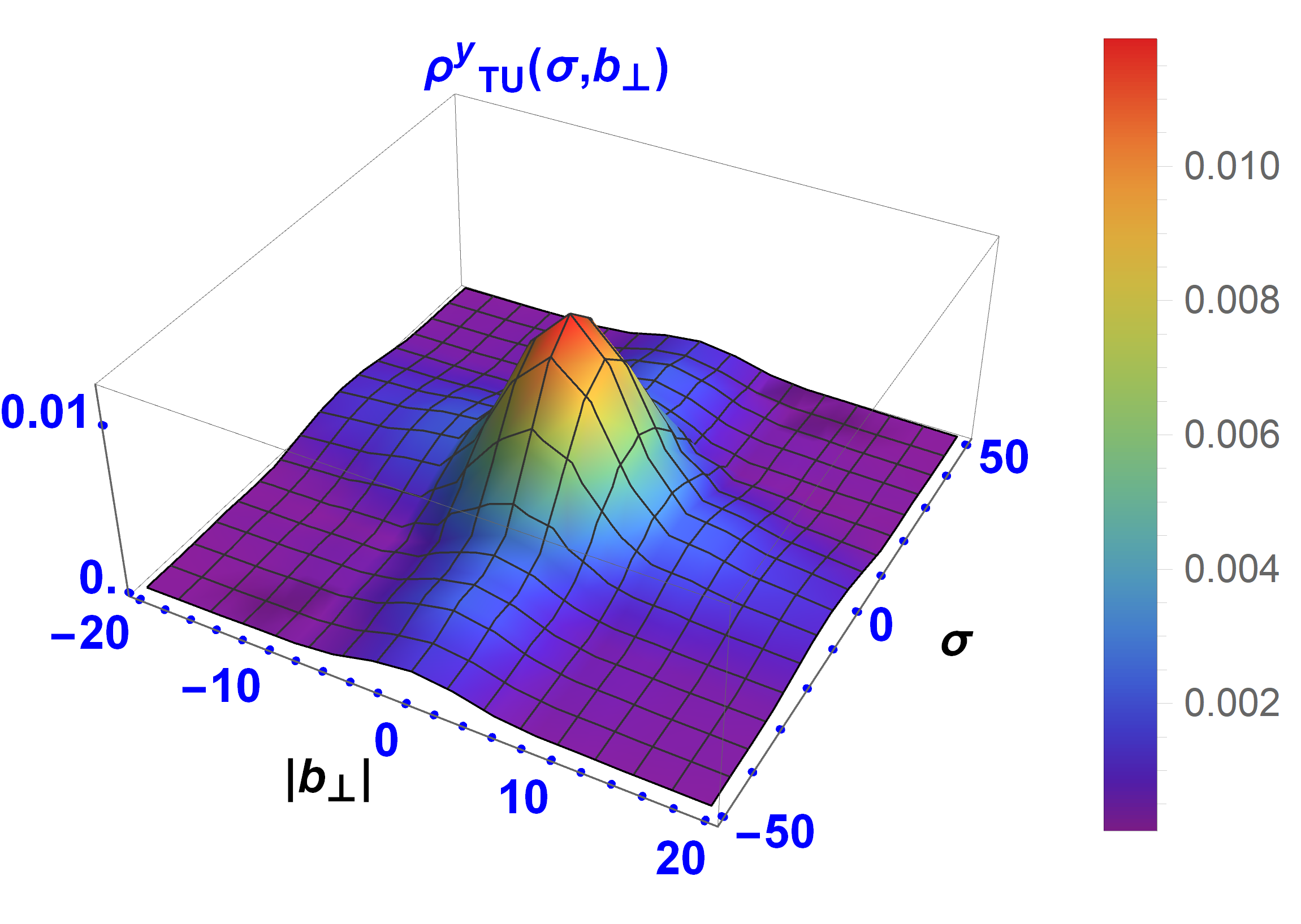}\\
\small{(c)}\includegraphics[width=8cm,height=5cm,clip]{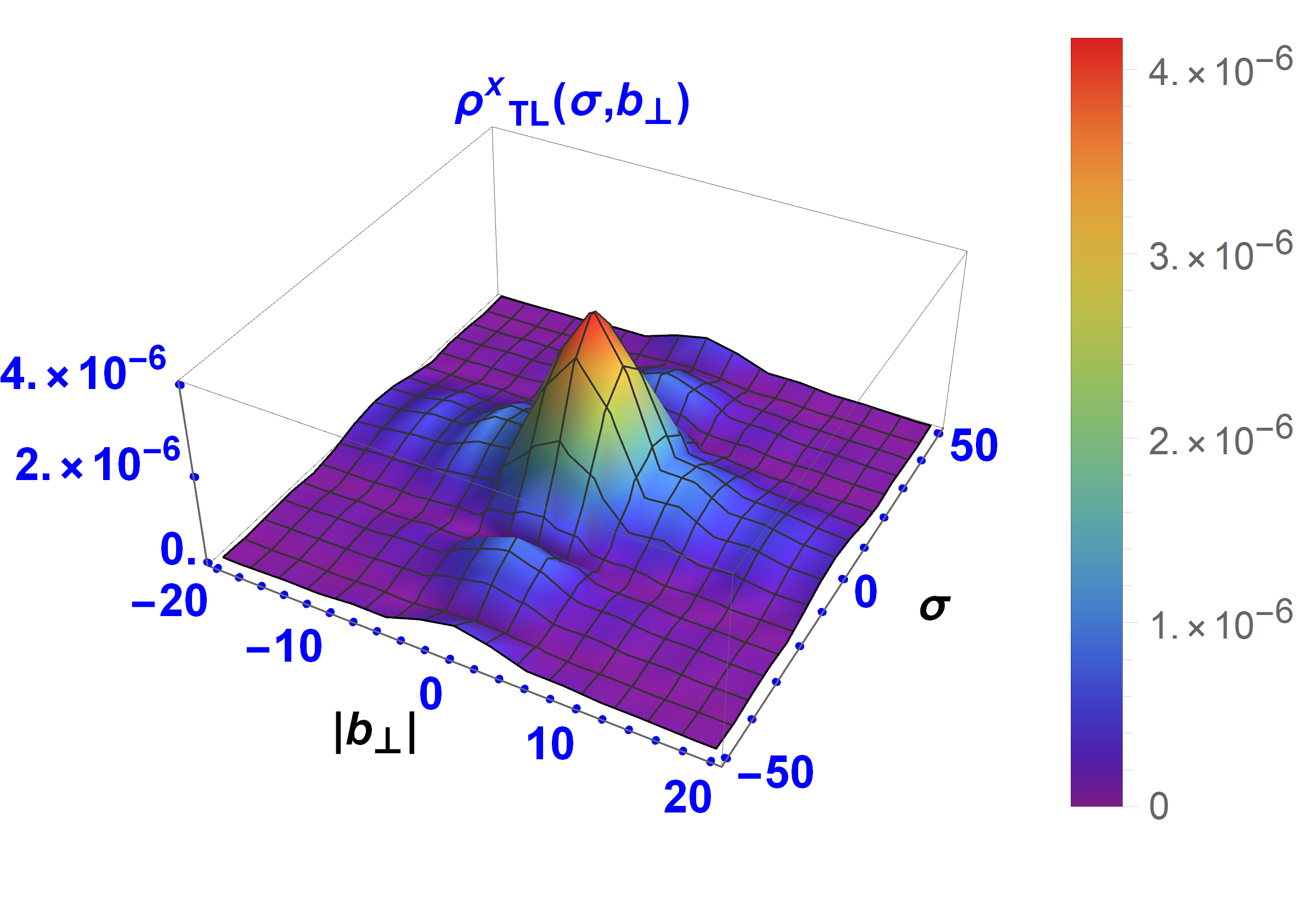}\\
\small{(d)}\includegraphics[width=8cm,height=5cm,clip]{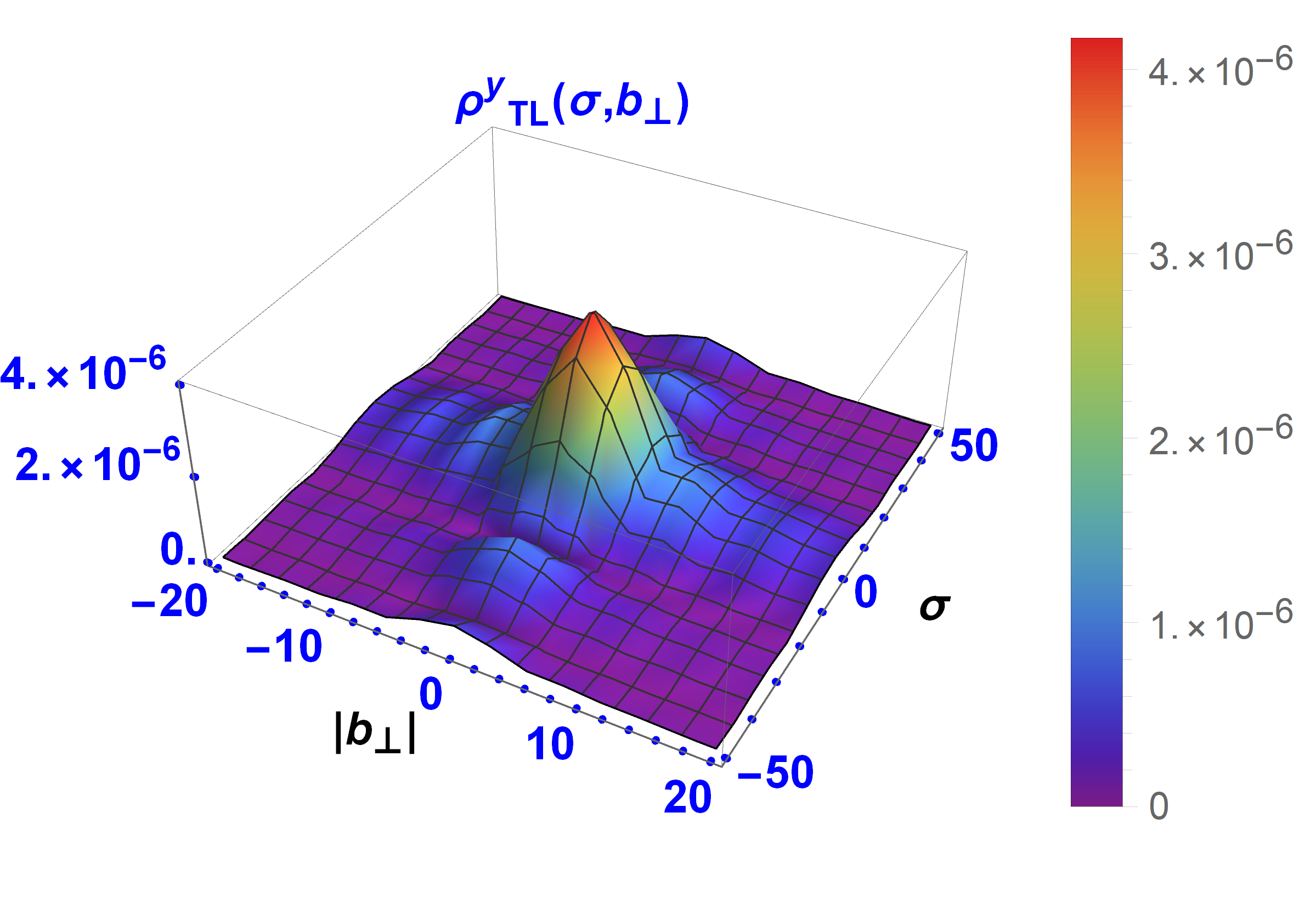}
\end{minipage}
\caption{\label{FigWigtrans}  Quark Wigner distributions in the transversely polarized target for various quark polarization states: (a) unpolarized, (b) longitudinally polarized, and (c), (d) transversely polarized.}
\end{figure}

%%%%%%%%%%%------------section---------%%%%%%%%%

\section{Result and Discussion}\label{RD}
In this section, the quark Wigner distributions, as defined in Eqs.(\ref{eqn.rhuu}-\ref{eqn.rhTL}), are illustrated in three-dimensional coordinate space ($|b_\perp|-\sigma$). In all plots, we fixed $x=0.3$, and integrated out the quark's transverse momentum over the range $(0\rightarrow0.5)$ GeV.

Figs. \ref{FigWigun}, \ref{FigWiglong}, and \ref{FigWigtrans} illustrate the Wigner distributions of quark with different polarization in unpolarized, longitudinally polarized and transversely polarized target state. 
A similar approach can be used to determine the quark spatial distribution inside other hadrons. The observations from the plots indicate that the quark distribution is primarily concentrated around \( b_\perp = 0 \) and \( \sigma = 0 \), rapidly decreasing as \( b_\perp \) and \( \sigma \) increase. The distribution exhibits symmetry in both \( \sigma \)-space and \( b_\perp \)-space, implying that the probability of finding a quark with a fixed $x$  at a longitudinal distance \( \sigma \) is the same on both sides of the origin. Likewise, the probability of finding a quark with a fixed $x$ at a transverse distance \( b_\perp \) from the center is also symmetric.  

The polarization of either the target state or the quark has a minimal impact on the overall nature of the distribution. However, it influences the magnitude of peaks and troughs, enhancing their contrast and making the distribution sharper. Notably, there are regions in both \( b_\perp \)-space and \( \sigma \)-space where the quark probability density is zero, and these regions are symmetrically distributed around the origin. This behavior resembles atomic orbitals, where certain regions have a higher probability of occupation than others, suggesting a form of spatial quantization around the center of the target. This quantization effect becomes more pronounced for some particular polarization configurations of the target and struck quark.

\section{Conclusion}
%In this paper, we have shown the quark Wigner distributions as functions of both longitudinal and transverse impact parameters, keeping the longitudinal momentum fraction constant. Moreover, we have visualized these distributions through plots, and it is significant to point out that the location of the quark probability density within the hadron depends on the polarization states of both the target and the quark. We observed that, for all combinations of quark and target polarizations, the quark's probability density is highest at the center of the hadron. 

We have computed the quark Wigner distributions in a frame-independent three-dimensional position space within the dressed quark model, considering various polarization configurations of both the quark and the target state. These distributions were visualized through numerical plots, revealing a clear dependence on the polarization states of both constituents. The Wigner distributions exhibit peak intensity near the center and gradually diminish outward, featuring oscillatory patterns with symmetric maxima and minima. This behavior is reminiscent of spatial quantization observed in atomic orbitals, where discrete structures emerge around the center of the atom.

A natural extension of this work would involve calculating the quark Wigner distributions in more complex systems, such as hadrons and mesons, to gain deeper insight into their internal partonic structure. Additionally, investigating the gluon Wigner distributions in a similar frame-independent three-dimensional coordinate space would be an interesting direction for future research.

% We obtained the quark Wigner distribution in the frame-independent 3-dimensional position space in the dressed quark model for different polarizatio of quark and target. Moreover, we visualized these distribution through plots and found that the quark Wigner distribution depends on the polarization states of both : the quark and the target. The distribution are maximum neat the centre and then reduces gradually with some maxima and minima symmetric around the centre as we move away from the centre. This behaviour resembles the space quantization similar to atomic orbital around the centre of atom.  Future work in this direction would be to get quark Wigner distribution inside different targets like hadrons and Mesons. It would also be intresting to obtain the gluon Wigner distribution in the frame-independent 3-dimensional position space. 

% However, for an unpolarized quark with a transversely polarized target in the x-direction, this probability density becomes minimal at the hadron’s center. 

%%%%%%%%%%%%%---------Reference------%%%%%%%%%%%

\bibliographystyle{elsarticle-num}
\bibliography{Ref.bib}
\end{document}